\documentclass[11pt,a4paper]{article}
\usepackage[margin=2.4cm]{geometry}
\usepackage[T1]{fontenc}
\usepackage[utf8]{inputenc}
\usepackage{microtype}
\usepackage{booktabs}
\usepackage{amsmath}
\usepackage{url}
\usepackage[hidelinks]{hyperref}
\usepackage{titlesec}
\usepackage{tikz}
\usepackage{pgfplots}
\pgfplotsset{compat=1.18}
\usetikzlibrary{positioning,arrows.meta,fit}
\usepackage{caption}
\titleformat{\section}{\normalfont\large\bfseries}{\thesection}{0.6em}{}
\titleformat{\subsection}{\normalfont\normalsize\bfseries}{\thesubsection}{0.6em}{}
\title{\vspace{-1.2cm}\textbf{Comparability in the public olfactory record:\\volatile measurement, human perception,\\and the join between them}}
\author{Fabio Rovai\\\small The Tesseract Academy\\\small \texttt{fabio@thetesseractacademy.com}}
\date{\small 30 August 2026}

\begin{document}
\maketitle
\vspace{-0.6cm}

\begin{abstract}
\noindent
Research data infrastructure measures metadata \emph{completeness}, the share of
fields populated. This is the wrong quantity: a populated field is not a joinable
field. We measure \emph{comparability} instead: whether two
records can be used together, and for what.

Working from 180{,}877 records across eight public sources, we adapt measurement
invariance from psychometrics into a four-rung ladder, each rung
established by its own test and licensing one specific comparison. Over all
167{,}331 pairs of gas chromatography analyses in a public
metabolomics repository, 100\% reach the bottom rung, 58.85\% the second,
0.072\% the third, and \textbf{exactly two pairs} reach the rung that licenses
comparing values directly, an upper bound under a generous test.

The deficit is recoverable: the missing values survive in prose. We formalise
crosswalking as five operations ordered by reliability, each with a
characteristic failure mode. The temperature programme that determines retention is
structured in none of a second repository's 355 gas chromatography studies yet
written in prose in 78.6\%; we recover 206 ordered
programmes comprising 529 steps, audited over three rounds.

The perceptual record fails differently: its datasets do not share a language.
Among eight that declare they measure odour character in humans, 42.9\% of pairs
share no descriptor, and two studies rating the same molecules with the same word
agree at mean correlation 0.31. Joining volatility measurement to human percept
across the entire public record yields \textbf{132 molecules}.

Across seismology, meteorology, metrology and chemistry, a field is populated
when the primary consumer cannot complete the primary task without it: the same
optional, unvalidated field sits at 100\% in one discipline and near zero in
another. Resources should publish a conformance ledger, reporting what
is missing as a tracked quantity; we release one, with 48{,}356 triples.

\end{abstract}

\section{Introduction}

The standard diagnostic applied to a scientific data repository is a completeness
audit. Fields are enumerated, populated values are counted, and a percentage is
reported. Improvement programmes then target the percentage.

This measures the wrong thing. A field can be populated and useless. In the corpus
examined below, the collision energy of a mass spectrum is recorded on 84.2\% of
records and is written in 1{,}023 distinct ways, so joining on it exactly retains
9.3\% of the comparisons that a normalised join retains. Completeness reports
84.2\%. The usable figure is 9.3\%.

What a consumer needs to know is not whether a field is filled but whether two
records can be used together, and for what. That property is relational rather
than intrinsic, it is graded rather than binary, and, we argue, it is computable.

\subsection*{Scope}

The measurements reported here draw on 180{,}877 records from eight public
repositories, spectral archives and standards bodies, none of which were collected
for this work.

\begin{center}
\small
\begin{tabular}{@{}llr@{}}
\toprule
\textbf{Source} & \textbf{What was examined} & \textbf{Records} \\
\midrule
MassBank \cite{horai2010} & Reference spectra parsed & 137{,}030 \\
Pyrfume \cite{hamel2024} & Molecules in the master table & 10{,}299 \\
MetaboLights & Assay-file method columns censused & 8{,}203 \\
Pyrfume & Molecules carrying a percept & 5{,}872 \\
PubChem \cite{pubchem} & Identifier resolutions & 5{,}837 \\
BIPM KCDB & Capability records field-censused & 4{,}000 \\
MetaboLights \cite{haug2013} & Study records harvested & 3{,}338 \\
BIPM KCDB & Comparison identifiers attempted & 2{,}600 \\
FDSN & Seismic channels & 1{,}607 \\
WMO OSCAR & Station deployments & 1{,}444 \\
Metabolomics Workbench \cite{sud2016} & Study records harvested & 594 \\
Pyrfume & Archives, manifests and curation scripts & 53 \\
\midrule
\textbf{Total} & & \textbf{180{,}877} \\
\bottomrule
\end{tabular}
\end{center}

A ninth source, an all-discipline bibliographic index \cite{openalex}, supplied the 52{,}271 work
counts behind Section~\ref{sec:related} and is not included in that total because
those works were counted rather than read.

From this material we classify 167{,}331 pairs of analyses, test 3{,}358
descriptor pairs against a permutation null, recover 206 ordered thermal
programmes and 66 retention index reference series from prose, and release 48{,}356
triples.

Our contributions are as follows. We import a graded comparability framework from
psychometrics and instantiate it for physical measurement (Section
\ref{sec:ladder}). We set out crosswalking and derivation as a method, with a
taxonomy of five operations ordered by reliability, an error model with its mitigations, and an
auditing procedure (Section \ref{sec:method}). We measure the
separation between the literatures that solved this problem independently, finding
twelve of fifteen term pairs sharing no work at all (Section \ref{sec:related}). We compute it over a complete public corpus and report the
distribution (Section \ref{sec:chem}). We demonstrate that the dominant deficit is
recoverable from unstructured text at scale, with audited precision (Section
\ref{sec:prose}). We measure the equivalent property in the perceptual record,
where the failure mode is different (Section \ref{sec:percept}). We quantify the
join between the two (Section \ref{sec:bridge}). We test, across four disciplines,
what actually causes a field to be populated (Section \ref{sec:why}). And we argue
for the conformance ledger as a publication object (Section \ref{sec:ledger}).

\section{Background}
\label{sec:related}

We did not invent a comparability framework. We imported one, and the reason for
importing rather than inventing is itself evidence that it should transfer.

The problem of combining measurements made under different conditions has been
solved independently in at least eight disciplines, each under its own name:
measurement invariance in psychometrics, calibration transfer in chemometrics,
retrospective harmonization in epidemiology \cite{fortier2017}, batch effect correction in genomics,
degrees of equivalence in metrology \cite{cipmmra}, test equating in educational testing,
retention indices in analytical chemistry, and domain adaptation in machine
learning. Three of them independently arrived at the same mechanism, namely
measuring something known alongside the unknown to place both on a common scale,
which chemometrics calls standardization samples \cite{wang1992}, educational
testing calls anchor items, chromatography calls a retention index series
\cite{kovats1958,vandendool1963} and genomics calls spike-in controls.

We adopt the psychometric formulation specifically, because it is the only one of
the eight that treats comparability as \emph{graded and separately testable}
rather than as a transformation to be applied. Configural invariance \cite{meredith1993} establishes
that the same construct is measured; metric invariance that the scales share
units; scalar invariance that they share an origin. Each is established by its own
test and licenses a specific claim. That structure transfers. The statistical
machinery that establishes those levels does not, and is not used here.

A second borrowing bears directly on Section~\ref{sec:chem}. Rohrschneider \cite{rohrschneider1966} and later
McReynolds \cite{mcreynolds1970} characterised a stationary phase by the retention
indices of a small set of probe solutes, so that phase equivalence becomes a
distance between vectors rather than a match between trade names. The corpus we
audit contains 107 distinct strings naming one chemically equivalent phase, so the
problem that work solved is live and the solution is unused.

\subsection{Measuring the separation}

The separation is measurable. Using a bibliographic index covering all
disciplines, we counted works whose title or abstract contains each term, and then
works containing each pair.

Six of the eight terms are specific enough to count without ambiguity. Their
corpora total 52{,}271 works. Of the fifteen pairs, \textbf{twelve share no
work at all}, and the twenty-eight co-occurrences that exist fall in three pairs:
calibration transfer with domain adaptation (18), batch effect correction with
domain adaptation (7), and measurement invariance with test equating (3).

The exceptions are as informative as the zeros. The two that connect to domain
adaptation are the two whose problem is stated as making a model built in one
place work in another, which is what domain adaptation is called in machine
learning. The third pair is two names from inside psychometrics. \textbf{Every
cross-disciplinary pair returns zero}, including measurement invariance against
calibration transfer, which are the same idea in two vocabularies.

This measures co-occurrence of terminology rather than citation, so it is a lower
bound on contact: a paper can cite another without naming its term. It also
depends on term choice, and we report the terms used so the count can be
reproduced or disputed.

\begin{figure}[t]
\centering
\begin{tikzpicture}[
  x={(0.92cm,0cm)}, y={(0.40cm,0.30cm)}, z={(0cm,0.92cm)},
  font=\scriptsize,
  plane/.style={fill=black!5, draw=black!25},
  nd/.style={circle, draw=black!70, fill=white, inner sep=1.4pt},
  as/.style={circle, draw=black!85, fill=black!85, inner sep=1.9pt},
  ed/.style={draw=black!45, -{Latex[length=1.3mm]}},
  lb/.style={font=\scriptsize, text=black!65}
]
% three planes, bottom to top
\foreach \z/\name/\sub in {
  0/{Observations}/{signals, six laboratories},
  2.1/{Assertions}/{dated, attributed, supersedable},
  4.2/{Entities}/{compounds}}{
  \fill[plane] (0,0,\z) -- (6,0,\z) -- (6,3,\z) -- (0,3,\z) -- cycle;
  \node[lb,anchor=west,font=\scriptsize\bfseries] at (6.4,0.2,\z) {\name};
  \node[lb,anchor=west,font=\scriptsize] at (6.4,-1.15,\z) {\sub};
}

% observation layer: signals from three laboratories
\node[nd] (s1) at (1.0,0.7,0) {}; \node[nd] (s2) at (2.7,1.9,0) {};
\node[nd] (s3) at (4.4,0.9,0) {}; \node[nd] (s4) at (3.4,2.6,0) {};

% assertion layer
\node[as] (a1) at (1.3,1.1,2.1) {}; \node[as] (a2) at (3.0,2.0,2.1) {};
\node[as] (a3) at (4.2,1.2,2.1) {}; \node[as] (a4) at (2.2,2.5,2.1) {};

% entity layer
\node[nd,minimum size=4pt] (c1) at (2.0,1.4,4.2) {}; \node[nd,minimum size=4pt] (c2) at (4.0,1.8,4.2) {};

% edges upward
\foreach \a/\b in {s1/a1, s2/a2, s3/a3, s4/a4} \draw[ed] (\a) -- (\b);
\foreach \a/\b in {a1/c1, a2/c1, a4/c1, a3/c2} \draw[ed] (\a) -- (\b);

% the point of the picture, placed below rather than beside the layer labels
\node[lb, anchor=north, text width=11cm, align=center] at (3.0,-2.6,0)
  {Every path from an observation to an entity passes through an assertion.
   An identification can therefore be withdrawn without disturbing what was
   measured.};
\end{tikzpicture}
\caption{The model has three layers and no shortcuts between them. Observations sit
below, entities above, and every path between the two passes through an assertion
that carries a date, an agent and a method. The structure is what makes the archive
reinterpretable: a later identification adds a node in the middle plane and
supersedes its predecessor, leaving both the signal beneath it and the compound
above it untouched.}
\label{fig:layers}
\end{figure}

\section{A graded measure of comparability}
\label{sec:ladder}

We instantiate the psychometric levels for gas chromatography. The logic of
separately tested, cumulative levels transfers; the factor-analytic apparatus that
establishes them does not, and is not used.

\begin{center}
\small
\begin{tabular}{@{}p{3.1cm}p{6.2cm}p{5.6cm}@{}}
\toprule
\textbf{Rung} & \textbf{Established when} & \textbf{Licenses} \\
\midrule
Configural & Same measurand, same technique & The records are about the same thing \\
Metric, scale & Same stationary phase family & Comparing orderings \\
Metric, calibrated & A retention index against a stated reference series, in both & Comparing indices \\
Scalar & Same run conditions & Comparing values directly \\
\bottomrule
\end{tabular}
\end{center}

The rungs are cumulative and defined as predicates over an unordered pair of
analyses $(a,b)$. Writing $m$, $\phi$, $\sigma$ and $\pi$ for the recorded
modality, stationary phase family, retention index reference series and thermal
programme, and $\bot$ for a value the record does not carry:
\begin{align*}
L_1(a,b) &\iff m(a)=m(b) \\
L_2(a,b) &\iff L_1(a,b) \wedge \phi(a)=\phi(b) \wedge \phi(a)\neq\bot \\
L_3(a,b) &\iff L_2(a,b) \wedge \sigma(a)\neq\bot \wedge \sigma(b)\neq\bot \\
L_4(a,b) &\iff L_3(a,b) \wedge \pi(a)=\pi(b) \wedge \pi(a)\neq\bot
\end{align*}
and the rung of a pair is $\max\{i : L_i(a,b)\}$. Assignment uses only what is
recorded: no imputation, and no inference from publication text at this stage,
which is what separates Section~\ref{sec:chem} from Section~\ref{sec:prose}.

Two consequences follow from the definition and are worth making explicit. The
predicates are decidable from the record alone, so the classification is
reproducible by anyone with the same corpus. And because every pair in the corpus
is classified rather than sampled, the counts in Section~\ref{sec:chem} are a
census and carry no sampling error; the uncertainty in them is definitional, in
$\phi$ and $\pi$, and is discussed there.

\section{Crosswalking and derivation}
\label{sec:method}

The measurements in this paper were not collected. Every one was derived from
material already published, by a sequence of crosswalks between representations
that were never designed to meet. We set the method out here as a method, because
it is the part that transfers, and because its error modes are specific enough to
be worth naming.

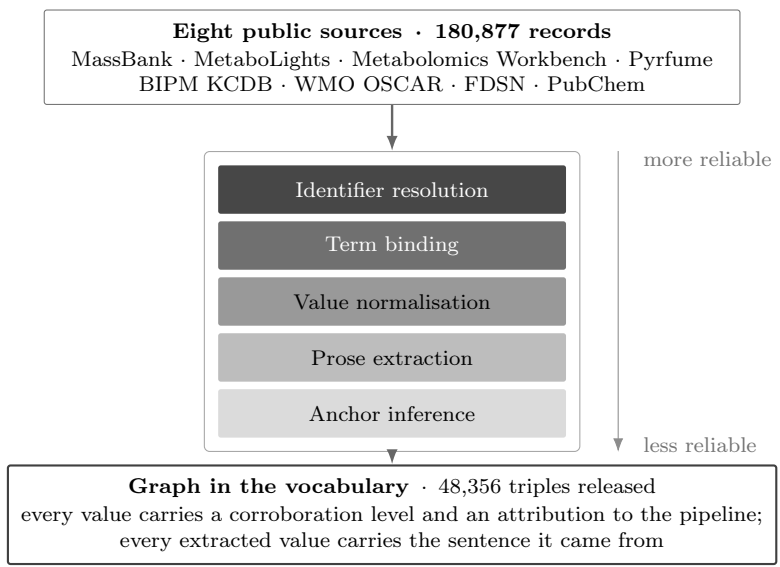
\begin{figure}[t]
\centering
\begin{tikzpicture}[
  font=\small, x=1cm, y=1cm,
  flat/.style={draw=black!55, rounded corners=1pt, align=center, font=\scriptsize,
               inner sep=5pt, minimum width=9.2cm},
  op/.style={rounded corners=1pt, inner sep=0pt, minimum width=4.6cm,
             minimum height=6.4mm, align=center, font=\scriptsize},
  ar/.style={-{Latex[length=2mm]}, draw=black!60, line width=0.9pt}
]
\node[flat] (src) at (0,0)
  {\textbf{Eight public sources \,\textperiodcentered\, 180{,}877 records}\\[1pt]
   MassBank \textperiodcentered{} MetaboLights \textperiodcentered{} Metabolomics Workbench \textperiodcentered{} Pyrfume\\
   BIPM KCDB \textperiodcentered{} WMO OSCAR \textperiodcentered{} FDSN \textperiodcentered{} PubChem};

\foreach \i/\shade/\name in {0/72/{Identifier resolution}, 1/56/{Term binding},
                             2/40/{Value normalisation}, 3/26/{Prose extraction},
                             4/14/{Anchor inference}}
{
  \node[op, fill=black!\shade, text=\ifnum\shade>45 white\else black\fi]
    (op\i) at (0,-1.75-0.74*\i) {\name};
}
\node[draw=black!35, rounded corners=2pt, fit=(op0)(op4), inner sep=5pt] (stack) {};

\node[flat, draw=black!75, thick] (g) at (0,-6.05)
  {\textbf{Graph in the vocabulary} \,\textperiodcentered\, 48{,}356 triples released\\[1pt]
   every value carries a corroboration level and an attribution to the pipeline;\\
   every extracted value carries the sentence it came from};

\draw[ar] (src.south) -- (stack.north);
\draw[ar] (stack.south) -- (g.north);

\draw[-{Latex[length=1.8mm]}, draw=black!45] ([xshift=5mm]stack.north east) -- ([xshift=5mm]stack.south east);
\node[font=\scriptsize, text=black!55, anchor=west] at ([xshift=7mm,yshift=-1mm]stack.north east) {more reliable};
\node[font=\scriptsize, text=black!55, anchor=west] at ([xshift=7mm,yshift=1mm]stack.south east) {less reliable};
\end{tikzpicture}
\caption{Derivation rather than collection. Public records pass through five
crosswalk operations, shaded by reliability. Reliability falls down the stack and
so does the strength of claim a derived value supports: an identifier resolution
is checkable by anyone with registry access, an anchor inference is a statistical
claim a later method may overturn. Nothing here was measured for this work.}
\label{fig:pipeline}
\end{figure}

\subsection{A taxonomy of crosswalk operations}

Joining two records made under different conventions is routinely described as
mapping, as though a single kind of thing were happening. Five distinct operations
occur in this work, and they differ by an order of magnitude in reliability.
Treating them alike is how derived corpora acquire confident errors.

\begin{center}
\small
\begin{tabular}{@{}p{3.1cm}p{3.3cm}p{3.9cm}p{4.2cm}@{}}
\toprule
\textbf{Operation} & \textbf{Mechanism} & \textbf{Characteristic failure} & \textbf{Observed here} \\
\midrule
Identifier resolution & Lookup in a governing registry & Silent absence: an unresolved record looks like an absent one & 5{,}837 of 5{,}872 molecules resolved; the remainder are absent from the bridge, not known to be unmeasured \\
\addlinespace
Term binding & Rule from a source string to a declared vocabulary & Invented terms: a value forced onto the nearest available concept & 2{,}106 spectra bound to a modality, 134{,}924 excluded and counted rather than bound \\
\addlinespace
Value normalisation & Canonical form for a free-text quantity & Over-merging: distinct things collapsed by a loose pattern & Exact join retains 197 comparisons, normalised join 1{,}724, an 8.7-fold recovery \\
\addlinespace
Prose extraction & Pattern over unstructured text, audited & Decoys and fabrication: a neighbouring value read as the target & 206 programmes and 66 reference series recovered; audits reported in Section~\ref{sec:prose} \\
\addlinespace
Anchor inference & Statistics over items measured by both parties & Multiple comparisons: the largest of many correlations is large regardless & 29 of 3{,}358 descriptor pairs survive a permutation null at 0.544 \\
\bottomrule
\end{tabular}
\end{center}

The ordering is deliberate. Reliability falls monotonically down the table, and so
does the strength of claim a derived value can support. An identifier resolution
is checkable by anyone with access to the registry. An anchor inference is a
statistical claim that a later method may overturn, and in our case establishes
only the dominant axis of a perceptual space rather than the vocabulary mapping it
was run to find.

\subsection{Provenance and grading of derived values}

It follows that a derived corpus cannot store its values as though they were
observations. Each carries a different warrant, and a consumer who cannot tell
them apart will use the weakest as though it were the strongest.

Every derived value released here therefore carries three things beyond its
content. It carries a \emph{corroboration level}, drawn from a closed set, which
states how far the value was verified. It carries an \emph{attribution} to the
extracting pipeline and a date, so it is a claim by us about what a source says
rather than a statement by the source's author. And where it came from text, it
carries the \emph{sentence it was extracted from}, so a reader can adjudicate the
extraction by reading rather than by rerunning anything.

That last property is the one we would most like to see adopted. An extraction at
scale is a large number of small claims, and an extraction that cannot show its
evidence is a way of manufacturing confident errors quickly.

\subsection{The error model}

Four of the five operations failed in the course of this work. The failures are
reported here rather than in a limitations paragraph because each has a specific
mitigation, and the mitigations are the method.

\textbf{Decoy capture.} Oven temperatures share their clauses with inlet,
injector, transfer line and ion source temperatures, at similar values. An
extractor that takes the first temperature after an anchor returns a programme
beginning at the injector. Mitigation: reject any candidate governed by a decoy
term within a bounded window, and verify on cases where a decoy sits in the same
sentence as the target.

\textbf{Fabrication from adjacency.} A study whose final temperature is followed
by \emph{``the total run time was 50 min''} yielded a fifty minute hold that never
happened. Mitigation: exclude durations governed by run-time and ramp-duration
language, which on introduction immediately caught a second instance.

\textbf{Substring collapse.} A pattern containing \texttt{5ms}, bounded only on
the right, matches inside \texttt{DB-35ms}, a chemically different phase. This
class of fault occurred three times in this work, twice more than we would have
predicted, and each time produced a plausible number. Mitigation: bound both ends
of every pattern that reads a chemical identifier, and treat an identifier as a
token rather than a substring.

\textbf{Vacuous success.} A verdict written as \texttt{not transfer or mean(...)
> 0.58} is satisfied by an empty list. Combined with a join that returned nothing,
because the two corpora key their tables by different identifier systems, it
asserted that a result survived a change of corpus at the moment it had failed to
test that. Mitigation: a test that cannot fail is not a test, and every aggregate
check should fail closed on an empty input. The same fault appeared independently
in our conformance layers and in a completeness table that printed zeroes computed
from a failed network fetch.

\subsection{Auditing an extraction}

Precision on an extraction at scale cannot be assumed and cannot be established by
inspecting the output distribution, which looks reasonable whether or not the
values are right. We draw a random sample, print each extracted value beside the
sentence it came from, and adjudicate by reading.

Two rules follow. An extractor that changes must be re-audited
from a fresh draw, because a previous verdict measures the previous extractor; the
thermal programme extractor was audited three times for this reason. And an
extraction whose audited precision is poor should be withdrawn rather than
qualified: our internal-standard extractor reported candidates in 138 studies and
returned four compounds in a random twelve, so no yield is reported for it and
nothing it produced is in the release.

\subsection{Precision of small audits}

Audits of this size are cheap and correspondingly weak. Wilson intervals \cite{wilson1927} at 95\%
on every precision reported in this paper:

\begin{center}
\small
\begin{tabular}{@{}llrl@{}}
\toprule
\textbf{Extractor} & \textbf{Audit} & \textbf{Point} & \textbf{95\% interval} \\
\midrule
Reference series & 12 of 12 & 100\% & [76\%, 100\%] \\
Thermal programme, final & 10 of 14 & 71\% & [45\%, 88\%] \\
Thermal programme, after tightening & 11 of 12 & 92\% & [65\%, 99\%] \\
Thermal programme, first version & 7 of 12 & 58\% & [32\%, 81\%] \\
Internal standard, withdrawn & 4 of 12 & 33\% & [14\%, 61\%] \\
\bottomrule
\end{tabular}
\end{center}

The intervals are wide and bind in both directions. A perfect twelve is consistent
with a true precision of 76\%, so no extraction here is exact. The withdrawn
extractor's interval reaches 61\%, still far below usable, so that decision does
not rest on the small sample.

What the audits establish reliably is the ordering, and the ordering is what
governed the work: the first thermal extractor was worse than the second, the
second was better than the first, and the internal-standard extractor was worse
than both by a margin no sample of twelve could obscure. A larger audit would
narrow these intervals and is the obvious next thing to run. It would not change
which extractors shipped.

\section{The chemical record}
\label{sec:chem}

\begin{figure}[t]
\centering
\begin{tikzpicture}[x=1cm,y=1cm]
\definecolor{keep}{gray}{0.30}
\definecolor{lost}{gray}{0.82}
% rung, y, kept fraction of full width, label, count
\foreach \y/\w/\name/\n in {
  0/12.0/{Configural}/{167{,}331},
  -1.5/7.06/{Metric, scale}/{98{,}479},
  -3.0/0.30/{Metric, calibrated}/{120},
  -4.5/0.12/{Scalar}/{2}}
{
  \fill[lost] (0,\y) rectangle (12,\y+0.62);
  \fill[keep] (0,\y) rectangle (\w,\y+0.62);
  \node[anchor=east,font=\small] at (-0.25,\y+0.31) {\name};
  \node[anchor=west,font=\small\bfseries] at (12.25,\y+0.31) {\n};
}
% loss annotations between rungs
\node[anchor=west,font=\scriptsize\itshape,text=black!55] at (0.2,-0.72)
  {$-68{,}852$ pairs: the two analyses used different stationary phases};
\node[anchor=west,font=\scriptsize\itshape,text=black!55] at (0.2,-2.22)
  {$-98{,}359$ pairs: at least one states no retention index reference series};
\node[anchor=west,font=\scriptsize\itshape,text=black!55] at (0.2,-3.72)
  {$-118$ pairs: the two ran different thermal programmes};
\end{tikzpicture}
\caption{Where comparability is lost, and to what. Bar length is proportional to
the number of pairs reaching each rung, on a linear scale, so the third and fourth
rungs are drawn at a visible minimum rather than at true width. The second loss is
the one that governs the result: it removes 99.9\% of what survived the first, and
its cause is a single unrecorded field.}
\label{fig:ladder}
\end{figure}
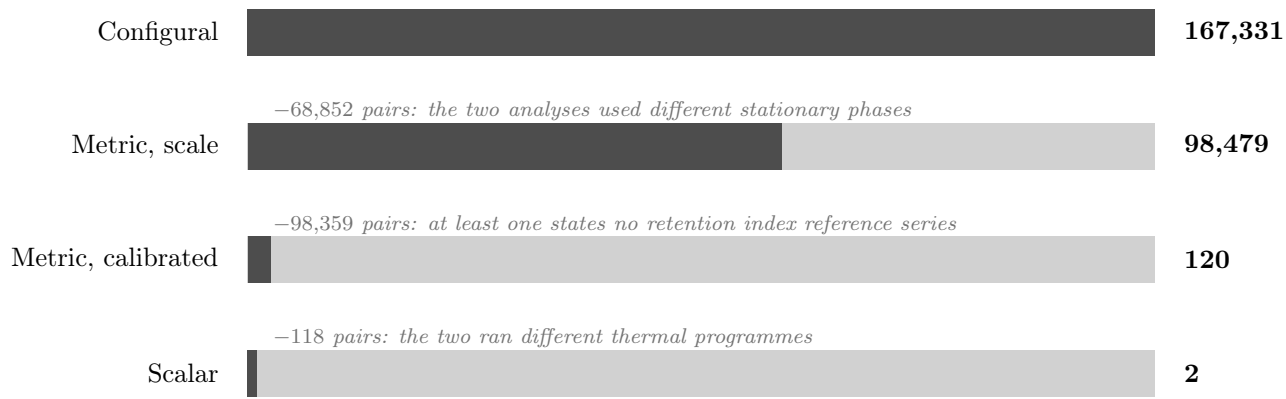

We classified every unordered pair of the 579 gas chromatography analyses in the
Metabolomics Workbench corpus, giving 167{,}331 pairs. Figure~\ref{fig:ladder}
shows the result.

\begin{center}
\small
\begin{tabular}{@{}lrr@{}}
\toprule
\textbf{Rung} & \textbf{Pairs at or above} & \textbf{Share} \\
\midrule
Configural & 167{,}331 & 100\% \\
Metric, scale & 98{,}479 & 58.85\% \\
Metric, calibrated & 120 & 0.072\% \\
Scalar & \textbf{2} & \textbf{0.001\%} \\
\bottomrule
\end{tabular}
\end{center}

Two pairs in one hundred and sixty-seven thousand reach the rung that licenses
comparing values. Programme equality is tested on the ordered sequence of numbers
in a free-text field, which treats two programmes as identical when one states a
run time the other omits, so the top rung is an upper bound.

The fall from the second rung to the third is a factor of approximately 800 and
has a single cause. The retention index reference series is stated for 19 of 579
analyses (3.3\%), and the convention by which the index was computed for
\textbf{one analysis}. Kováts interpolates logarithmically for isothermal
operation \cite{kovats1958}; van den Dool and Kratz interpolate linearly for
temperature-programmed operation \cite{vandendool1963}. Which is correct depends
on how the oven ran, and a bare retention index does not say.

\section{Recovery from prose}
\label{sec:prose}

\begin{figure}[t]
\centering
\begin{tikzpicture}[x=0.105cm,y=1cm]
\draw[->,black!45] (0,-0.55) -- (85,-0.55);
\foreach \x in {0,20,40,60,80} \node[font=\scriptsize,text=black!55] at (\x,-0.85) {\x\%};
\foreach \y/\lab/\a/\b/\c in {
  0/{Thermal programme}/0/58.0/78.6,
  1.15/{Retention index series}/3.3/18.6/80.0}
{
  \node[anchor=east,font=\small] at (-3,\y) {\lab};
  \draw[black!30,line width=1.1pt] (\a,\y) -- (\c,\y);
  \draw[black!62,line width=2.6pt] (\a,\y) -- (\b,\y);
  \fill[black!30] (\c,\y) circle (3.4pt);
  \fill[black!62] (\b,\y) circle (3.6pt);
  \fill[black] (\a,\y) circle (2.6pt);
  \node[font=\scriptsize,anchor=south] at (\b,\y+0.10) {\b\%};
  \node[font=\scriptsize,anchor=south,text=black!55] at (\c,\y+0.10) {\c\%};
}
\begin{scope}[shift={(0,2.15)}]
  \fill[black] (0,0) circle (2.6pt); \node[anchor=west,font=\scriptsize] at (2,0) {in a structured field};
  \fill[black!62] (34,0) circle (3.6pt); \node[anchor=west,font=\scriptsize] at (36,0) {recovered from prose};
  \fill[black!30] (68,0) circle (3.4pt); \node[anchor=west,font=\scriptsize] at (70,0) {mentioned in prose};
\end{scope}
\end{tikzpicture}
\caption{What the schema holds, what the text holds, and how much of the gap
extraction closes. The pale segment is the ceiling set by how often the value is
discussed at all; the dark segment is what was recovered and audited. The thermal
programme begins at zero because no structured field for it exists in that
corpus.}
\label{fig:recovery}
\end{figure}
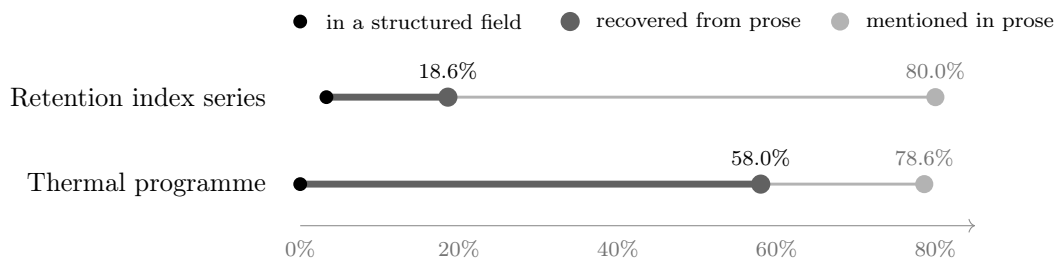

The values are not absent from the record. They are absent from the schema
(Figure~\ref{fig:recovery}).

In the MetaboLights corpus, the temperature programme is recorded in a structured
field for \textbf{none} of the 355 gas chromatography studies, and is described in
protocol prose in 279 of them (78.6\%). We built an extractor that reads the
programme as an ordered token stream of temperatures, ramp rates and durations,
excluding candidates governed by decoy terms such as inlet, injector, transfer
line and ion source, which carry temperatures in the same clauses at similar
values.

We recovered \textbf{206 ordered programmes} (58.0\% of gas chromatography
studies) comprising \textbf{529 temperature steps}, of which 370 carry a hold
time, giving 135 distinct programmes.

The extractor was audited three times, each on a fresh random draw, because it
changed twice and reusing a verdict would have measured the previous version. A
first implementation using independent regular expressions scored 7 of 12, with
all five failures multi-ramp: relative order between independent matches is lost,
producing rates paired with the following clause's target and, in one case, two
named methods spliced into a single programme. After replacement by the ordered
token walk the audit scored 11 of 12, then 10 of 14 with one genuinely wrong value
in which \emph{``the total run time was 50 min''} was read as a hold. Run times,
analysis times and ramp durations are now excluded explicitly. The remaining
failure mode omits a hold rather than inventing one, which is the safe direction:
an omitted value is a visible gap and a fabricated one is indistinguishable from a
measurement.

The retention index reference series shows the same pattern. It is a structured
field for 3.3\% of analyses, appears in protocol prose in 284 of 355 studies, and
is recoverable as a normalised carbon series for \textbf{66 studies (18.6\%)}, an
increase of 5.6-fold over the structured baseline. Twelve recovered series were
drawn at random and adjudicated against their evidence windows: twelve of twelve
were correct.

A companion attempt to recover internal standard identities from the same prose
was \textbf{withdrawn}. It reported candidates in 138 studies, but a random audit
of twelve returned four compounds and eight fragments of English such as
\emph{``and vortexed''} and \emph{``see below''}. A precision near one third is
not a measurement. Extracting a compound name requires a chemical entity
recogniser and a registry lookup, not a regular expression.

\section{The perceptual record}
\label{sec:percept}

We harvested 53 archives of the public human psychophysics corpus and asked the
same question. The failure mode is different and worse.

Restricting to the eight datasets that declare in their own manifests that they
measure odour character in humans, \textbf{42.9\% of pairs share no descriptor at
all}, with mean pairwise Jaccard overlap of 0.061.

\begin{figure}[t]
\centering
\begin{tikzpicture}[x=0.86cm,y=0.86cm]
\fill[black!8] (0,0) rectangle (1,-1);
\fill[black!26] (1,0) rectangle (2,-1);
\fill[black!13] (2,0) rectangle (3,-1);
\fill[black!90] (3,0) rectangle (4,-1);
\node[font=\tiny,text=white] at (3.5,-0.5) {0.37};
\fill[black!6] (4,0) rectangle (5,-1);
\node[font=\tiny,text=black!45] at (4.5,-0.5) {0};
\fill[black!78] (5,0) rectangle (6,-1);
\node[font=\tiny,text=white] at (5.5,-0.5) {0.32};
\fill[black!8] (6,0) rectangle (7,-1);
\fill[black!6] (7,0) rectangle (8,-1);
\node[font=\tiny,text=black!45] at (7.5,-0.5) {0};
\fill[black!26] (0,-1) rectangle (1,-2);
\fill[black!8] (1,-1) rectangle (2,-2);
\fill[black!12] (2,-1) rectangle (3,-2);
\fill[black!38] (3,-1) rectangle (4,-2);
\node[font=\tiny,text=black] at (3.5,-1.5) {0.14};
\fill[black!6] (4,-1) rectangle (5,-2);
\node[font=\tiny,text=black!45] at (4.5,-1.5) {0};
\fill[black!35] (5,-1) rectangle (6,-2);
\node[font=\tiny,text=black] at (5.5,-1.5) {0.13};
\fill[black!9] (6,-1) rectangle (7,-2);
\fill[black!6] (7,-1) rectangle (8,-2);
\node[font=\tiny,text=black!45] at (7.5,-1.5) {0};
\fill[black!13] (0,-2) rectangle (1,-3);
\fill[black!12] (1,-2) rectangle (2,-3);
\fill[black!8] (2,-2) rectangle (3,-3);
\fill[black!15] (3,-2) rectangle (4,-3);
\fill[black!6] (4,-2) rectangle (5,-3);
\node[font=\tiny,text=black!45] at (4.5,-2.5) {0};
\fill[black!11] (5,-2) rectangle (6,-3);
\fill[black!29] (6,-2) rectangle (7,-3);
\node[font=\tiny,text=black] at (6.5,-2.5) {0.10};
\fill[black!6] (7,-2) rectangle (8,-3);
\node[font=\tiny,text=black!45] at (7.5,-2.5) {0};
\fill[black!90] (0,-3) rectangle (1,-4);
\node[font=\tiny,text=white] at (0.5,-3.5) {0.37};
\fill[black!38] (1,-3) rectangle (2,-4);
\node[font=\tiny,text=black] at (1.5,-3.5) {0.14};
\fill[black!15] (2,-3) rectangle (3,-4);
\fill[black!8] (3,-3) rectangle (4,-4);
\fill[black!6] (4,-3) rectangle (5,-4);
\node[font=\tiny,text=black!45] at (4.5,-3.5) {0};
\fill[black!89] (5,-3) rectangle (6,-4);
\node[font=\tiny,text=white] at (5.5,-3.5) {0.37};
\fill[black!13] (6,-3) rectangle (7,-4);
\fill[black!6] (7,-3) rectangle (8,-4);
\node[font=\tiny,text=black!45] at (7.5,-3.5) {0};
\fill[black!6] (0,-4) rectangle (1,-5);
\node[font=\tiny,text=black!45] at (0.5,-4.5) {0};
\fill[black!6] (1,-4) rectangle (2,-5);
\node[font=\tiny,text=black!45] at (1.5,-4.5) {0};
\fill[black!6] (2,-4) rectangle (3,-5);
\node[font=\tiny,text=black!45] at (2.5,-4.5) {0};
\fill[black!6] (3,-4) rectangle (4,-5);
\node[font=\tiny,text=black!45] at (3.5,-4.5) {0};
\fill[black!8] (4,-4) rectangle (5,-5);
\fill[black!6] (5,-4) rectangle (6,-5);
\node[font=\tiny,text=black!45] at (5.5,-4.5) {0};
\fill[black!6] (6,-4) rectangle (7,-5);
\node[font=\tiny,text=black!45] at (6.5,-4.5) {0};
\fill[black!6] (7,-4) rectangle (8,-5);
\node[font=\tiny,text=black!45] at (7.5,-4.5) {0};
\fill[black!78] (0,-5) rectangle (1,-6);
\node[font=\tiny,text=white] at (0.5,-5.5) {0.32};
\fill[black!35] (1,-5) rectangle (2,-6);
\node[font=\tiny,text=black] at (1.5,-5.5) {0.13};
\fill[black!11] (2,-5) rectangle (3,-6);
\fill[black!89] (3,-5) rectangle (4,-6);
\node[font=\tiny,text=white] at (3.5,-5.5) {0.37};
\fill[black!6] (4,-5) rectangle (5,-6);
\node[font=\tiny,text=black!45] at (4.5,-5.5) {0};
\fill[black!8] (5,-5) rectangle (6,-6);
\fill[black!10] (6,-5) rectangle (7,-6);
\fill[black!8] (7,-5) rectangle (8,-6);
\fill[black!8] (0,-6) rectangle (1,-7);
\fill[black!9] (1,-6) rectangle (2,-7);
\fill[black!29] (2,-6) rectangle (3,-7);
\node[font=\tiny,text=black] at (2.5,-6.5) {0.10};
\fill[black!13] (3,-6) rectangle (4,-7);
\fill[black!6] (4,-6) rectangle (5,-7);
\node[font=\tiny,text=black!45] at (4.5,-6.5) {0};
\fill[black!10] (5,-6) rectangle (6,-7);
\fill[black!8] (6,-6) rectangle (7,-7);
\fill[black!6] (7,-6) rectangle (8,-7);
\node[font=\tiny,text=black!45] at (7.5,-6.5) {0};
\fill[black!6] (0,-7) rectangle (1,-8);
\node[font=\tiny,text=black!45] at (0.5,-7.5) {0};
\fill[black!6] (1,-7) rectangle (2,-8);
\node[font=\tiny,text=black!45] at (1.5,-7.5) {0};
\fill[black!6] (2,-7) rectangle (3,-8);
\node[font=\tiny,text=black!45] at (2.5,-7.5) {0};
\fill[black!6] (3,-7) rectangle (4,-8);
\node[font=\tiny,text=black!45] at (3.5,-7.5) {0};
\fill[black!6] (4,-7) rectangle (5,-8);
\node[font=\tiny,text=black!45] at (4.5,-7.5) {0};
\fill[black!8] (5,-7) rectangle (6,-8);
\fill[black!6] (6,-7) rectangle (7,-8);
\node[font=\tiny,text=black!45] at (6.5,-7.5) {0};
\fill[black!8] (7,-7) rectangle (8,-8);
\node[anchor=east,font=\scriptsize] at (-0.15,-0.5) {arctander};
\node[anchor=east,font=\scriptsize] at (-0.15,-1.5) {dravnieks};
\node[anchor=east,font=\scriptsize] at (-0.15,-2.5) {keller};
\node[anchor=east,font=\scriptsize] at (-0.15,-3.5) {leffingwel};
\node[anchor=east,font=\scriptsize] at (-0.15,-4.5) {ravia};
\node[anchor=east,font=\scriptsize] at (-0.15,-5.5) {sigma};
\node[anchor=east,font=\scriptsize] at (-0.15,-6.5) {snitz};
\node[anchor=east,font=\scriptsize] at (-0.15,-7.5) {weiss};
\node[anchor=west,font=\scriptsize,rotate=55] at (0.5,0.15) {arctander};
\node[anchor=west,font=\scriptsize,rotate=55] at (1.5,0.15) {dravnieks};
\node[anchor=west,font=\scriptsize,rotate=55] at (2.5,0.15) {keller};
\node[anchor=west,font=\scriptsize,rotate=55] at (3.5,0.15) {leffingwel};
\node[anchor=west,font=\scriptsize,rotate=55] at (4.5,0.15) {ravia};
\node[anchor=west,font=\scriptsize,rotate=55] at (5.5,0.15) {sigma};
\node[anchor=west,font=\scriptsize,rotate=55] at (6.5,0.15) {snitz};
\node[anchor=west,font=\scriptsize,rotate=55] at (7.5,0.15) {weiss};
\draw[black!35] (0,0) rectangle (8,-8);
\end{tikzpicture}
\caption{Pairwise descriptor overlap, as Jaccard index, between the eight public
datasets that each declare in their own manifest that they measure odour character
in humans. Darker is more shared vocabulary and the scale is anchored at the
maximum observed value of 0.37. Twelve of the twenty-eight pairs share no
descriptor at all and are marked zero. The two darkest cells are perfumery
reference compilations rather than experiments; no pair of psychophysical studies
exceeds 0.10.}
\label{fig:overlap}
\end{figure}

Restriction to comparable measurands changes the figure substantially, and the
variable that permits the restriction is not in the data. An unrestricted
comparison across every archive whose descriptors can be read gives 86.7\%.
Restricted to human odour-character datasets it is 42.9\%. The difference is three
archives that are not human perception at all: two rodent studies and a receptor
assay in \emph{Harpegnathos}, an ant.

The stratifying variable is a declared measurement-type tag, shipped in a manifest
file alongside every archive and consumed by nothing. An analysis reading only the
data tables cannot distinguish a human panel from an insect receptor assay, and
will report their disjoint vocabularies as a finding about perception.

Robustness was checked before publication. Stemming reduced 460 terms to 453 and
left the disjoint pair count unchanged. Fuzzy matching at a 0.85 threshold rescued
two of ninety-four disjoint pairs in the unrestricted comparison. The vocabularies
differ in substance, not in spelling.

\subsection{Agreement between identical descriptors}

Two studies that rated an overlapping molecule set have, unintentionally, run the
anchor-item design. We tested this between a standardised odour character atlas
\cite{dravnieks1985} and a large modern psychophysics study \cite{keller2016}: 146
descriptors against 24, over 65 molecules rated by both, giving 3{,}358 descriptor
pairs. Because the maximum of thousands of correlations is large regardless, every
pair was tested against a null built by shuffling molecule labels, preserving both
marginals and destroying only the pairing. The 95th percentile of the maximum
spurious correlation is 0.544.

Twenty-nine pairs survive (0.86\%), and twenty-four of those are an atlas
descriptor against the modern study's pleasantness item, recovering the known
dominant axis of odour perception rather than any vocabulary mapping. Five of 24
modern descriptors receive any relation.

The sharper result concerns identical words. Of twelve descriptor pairs sharing a
word, two clear the threshold and the mean correlation is \textbf{0.31}. Sweet
against sweet is 0.59, which is the strongest relation found anywhere in the
analysis and serves as a validation of the method. Garlic against garlic is 0.27.
Musk against musky is 0.18.

Whether this reflects genuine disagreement or a concentration artefact
\textbf{cannot be determined from the record}. The atlas records concentration as
the word \texttt{high} or \texttt{low}; the modern study records dilutions. There
is no stated relation between a word and a dilution, and matching the closest
available correspondence leaves seven molecules, too few to compute a correlation.
Inspection of the curation script reveals that the atlas concentration field is
not a measurement at all: it is assigned by testing whether the molecule's
\emph{name} contains the substring \texttt{low}.

\subsection{The unused olfactometric reference}

We initially wrote that perception has no instrument-independent invariant. That
is incorrect. Environmental dynamic olfactometry anchors odour concentration to
n-butanol in nitrogen, expressing any odour in n-butanol mass equivalents, with
criteria for trueness and repeatability and an established proficiency testing
practice \cite{en13725}.

Across the 53 archives, the number mentioning olfactometry, odour units, the
standard, dynamic dilution, or n-butanol as a reference is \textbf{zero in every
case}. n-butanol appears as a stimulus molecule in 26 archives and as a reference
in none. Two communities measure the same organ and have no contact. The search
covers curated manifests and tables, not the underlying publications.

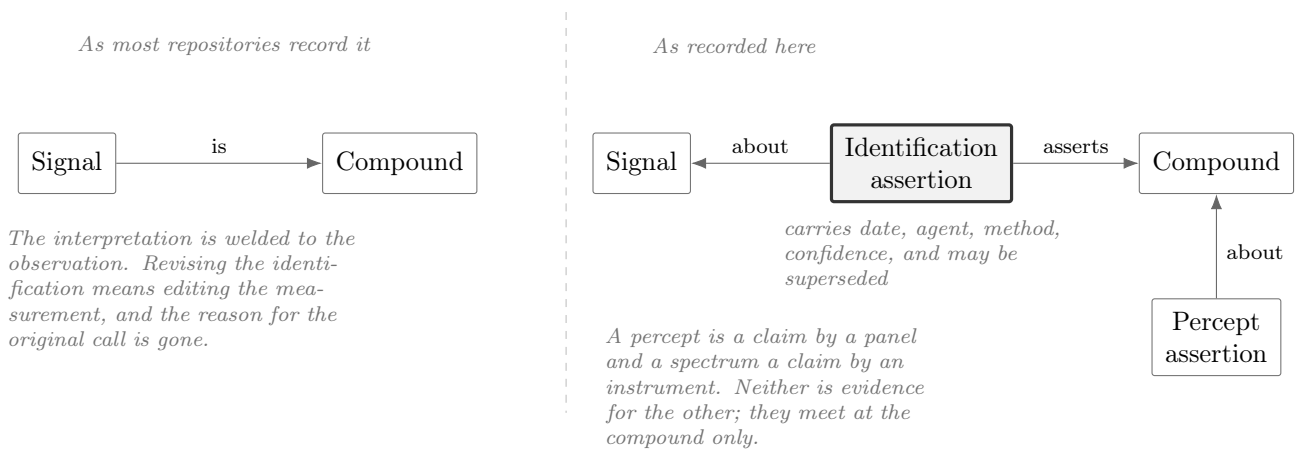
\begin{figure}[t]
\centering
\begin{tikzpicture}[
  font=\small,
  ent/.style={draw=black!55, rounded corners=1pt, inner sep=5pt, minimum height=8mm, align=center},
  asn/.style={draw=black!80, very thick, rounded corners=1pt, inner sep=5pt, fill=black!5, align=center},
  ann/.style={font=\scriptsize\itshape, text=black!55},
  ar/.style={-{Latex[length=2mm]}, draw=black!60}
]
% ---- naive
\node[ann,anchor=west] at (0,1.55) {As most repositories record it};
\node[ent] (s1) at (0,0) {Signal};
\node[ent] (c1) at (4.4,0) {Compound};
\draw[ar] (s1) -- node[above,font=\scriptsize]{is} (c1);
\node[ann,text width=4.6cm,anchor=north west] at (-0.9,-0.75)
  {The interpretation is welded to the observation. Revising the identification
   means editing the measurement, and the reason for the original call is gone.};

% ---- reified
\node[ann,anchor=west] at (7.6,1.55) {As recorded here};
\node[ent] (s2) at (7.6,0) {Signal};
\node[asn] (a2) at (11.3,0) {Identification\\assertion};
\node[ent] (c2) at (15.2,0) {Compound};
\draw[ar] (a2) -- node[above,font=\scriptsize]{about} (s2);
\draw[ar] (a2) -- node[above,font=\scriptsize]{asserts} (c2);
\node[ann,text width=3.6cm,anchor=north] at (11.3,-0.62)
  {carries date, agent, method, confidence, and may be superseded};
\node[ent,fill=white] (p2) at (15.2,-2.3) {Percept\\assertion};
\draw[ar] (p2) -- node[right,font=\scriptsize,pos=0.45]{about} (c2);
\node[ann,text width=4.2cm,anchor=north west] at (7.0,-2.05)
  {A percept is a claim by a panel and a spectrum a claim by an instrument.
   Neither is evidence for the other; they meet at the compound only.};
\draw[black!25,dashed] (6.6,2.0) -- (6.6,-3.3);
\end{tikzpicture}
\caption{The recording decision the vocabulary turns on. Binding a compound
directly to a signal, on the left, makes the observation and its interpretation
one object, so a later method cannot revise the identification without altering
the measurement and the basis of the original call is not retained. Separating
them, on the right, costs one node and makes identification a dated, attributed,
supersedable claim. The same separation keeps a percept and a spectrum joined at
the molecule and nowhere else.}
\label{fig:model}
\end{figure}

\section{The bridge}
\label{sec:bridge}

We built the join that an olfactory resource presupposes: one molecule carrying
both the laboratories that measured its volatility and the descriptors people
applied to it, with each side retaining its own provenance and neither treated as
evidence for the other.

Of 5{,}872 molecules carrying a percept, 5{,}837 resolved to a structural key
through a public compound registry. Of 137{,}030 spectra in the reference spectral
archive, 2{,}106 measure volatility by gas chromatography, covering 969 molecules.
The intersection is \textbf{132 molecules}, behind which sit 434 spectra from 6
laboratories and 147 distinct descriptors. \textbf{54} of the 132 are measured by
two or more laboratories, and for those the resource can state a cross-laboratory
agreement and a percept together.

A first version of this join reported 1{,}710 molecules by counting any mass
spectrum as a measurement. A liquid chromatography spectrum is a measurement of
the molecule but not of its volatility, and admitting it would let a compound
enter an olfactory resource on evidence unrelated to smell. The collapse from
1{,}710 to 132 is a result, not an artefact of strictness: the public record
connecting instrument behaviour to perception is approximately one hundred
molecules wide.

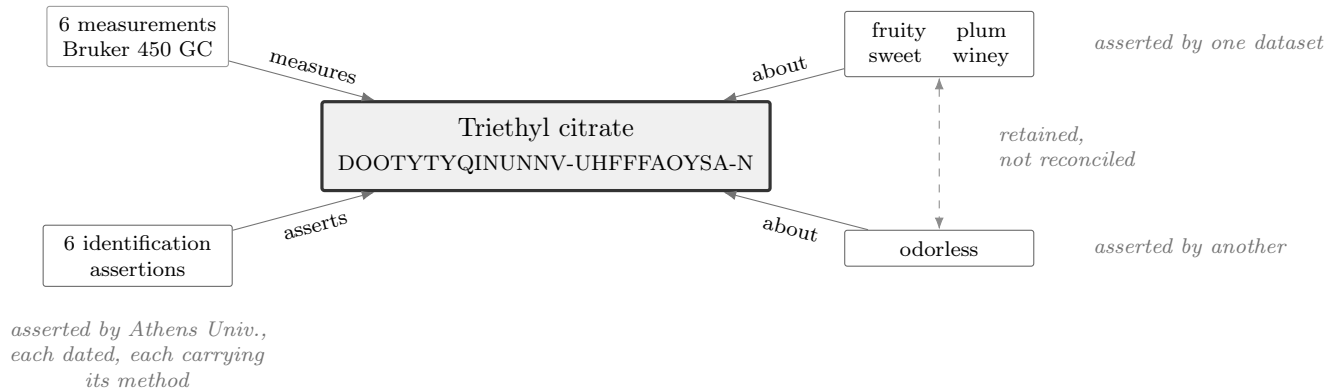
\begin{figure}[t]
\centering
\begin{tikzpicture}[
  font=\small, x=1cm, y=1cm,
  cmp/.style={draw=black!80, very thick, rounded corners=1pt, fill=black!6,
              inner sep=6pt, align=center},
  asn/.style={draw=black!60, rounded corners=1pt, inner sep=4pt, align=center,
              font=\scriptsize, minimum width=2.5cm},
  obs/.style={draw=black!40, rounded corners=1pt, inner sep=4pt, align=center,
              font=\scriptsize, minimum width=2.4cm},
  ar/.style={-{Latex[length=1.7mm]}, draw=black!55},
  ann/.style={font=\scriptsize\itshape, text=black!55}
]
\node[cmp] (c) at (0,0) {Triethyl citrate\\[1pt]\scriptsize DOOTYTYQINUNNV-UHFFFAOYSA-N};

% measurement side
\node[obs] (m) at (-5.4,1.45) {6 measurements\\Bruker 450 GC};
\node[asn] (i) at (-5.4,-1.45) {6 identification\\assertions};
\node[ann,text width=3.4cm,align=center] at (-5.4,-2.75)
  {asserted by Athens Univ.,\\each dated, each carrying\\its method};
\draw[ar] (m) -- node[above,sloped,font=\scriptsize,pos=0.55]{measures} (c);
\draw[ar] (i) -- node[below,sloped,font=\scriptsize,pos=0.55]{asserts} (c);

% percept side, showing the contradiction
\node[asn] (p1) at (5.2,1.35) {fruity \quad plum\\sweet \quad winey};
\node[asn] (p2) at (5.2,-1.35) {odorless};
\draw[ar] (p1) -- node[above,sloped,font=\scriptsize]{about} (c);
\draw[ar] (p2) -- node[below,sloped,font=\scriptsize]{about} (c);
\node[ann,anchor=west] at (7.1,1.35) {asserted by one dataset};
\node[ann,anchor=west] at (7.1,-1.35) {asserted by another};
\draw[black!45,dashed,{Latex[length=1.6mm]}-{Latex[length=1.6mm]}] (p1) -- (p2);
\node[ann,anchor=west,align=left] at (5.85,0.0) {retained,\\not reconciled};
\end{tikzpicture}
\caption{One molecule from the released bridge, drawn as it is recorded. Six
measurements from a single laboratory sit on the left, each with a separate dated
identification assertion. On the right, two public datasets disagree about whether
the compound has an odour at all. The model keeps both claims, attributed and
dated, because neither is evidence against the other and a resource that merged
them would destroy the disagreement rather than record it. Deciding between them
is a scientific act that someone must own, which is what the assertion node is
for.}
\label{fig:instance}
\end{figure}

\section{Determinants of field completeness}
\label{sec:why}

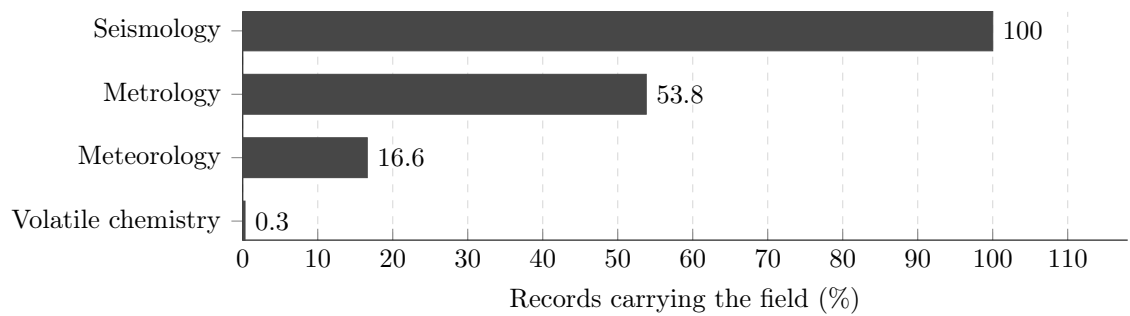
\begin{figure}[t]
\centering
\begin{tikzpicture}
\begin{axis}[
  width=0.82\textwidth, height=4.6cm,
  xbar, bar width=15pt, xmin=0, xmax=118,
  xlabel={Records carrying the field (\%)}, xlabel style={font=\small},
  tick label style={font=\small},
  symbolic y coords={{Volatile chemistry},{Meteorology},{Metrology},{Seismology}},
  ytick=data, y tick label style={font=\small},
  nodes near coords, nodes near coords style={font=\small\bfseries},
  xmajorgrids, grid style={dashed,gray!30}, axis lines*=left,
]
\addplot[fill=black!72, draw=black!72] coordinates
  {(100,{Seismology}) (53.8,{Metrology}) (16.6,{Meteorology}) (0.3,{Volatile chemistry})};
\end{axis}
\end{tikzpicture}
\caption{The field that makes a record interpretable, in four disciplines with
established machine-readable observation formats. Every one of these fields is
optional in its schema and none has a validator enforcing it, so neither the schema
nor enforcement explains the spread. The seismic instrument response, declared
\texttt{minOccurs="0"}, was measured at 1{,}607 of 1{,}607 channels on live data
centre holdings.}
\label{fig:why}
\end{figure}

We examined four disciplines with established machine-readable observation
formats, asking in each what the schema requires and what records contain
(Figure~\ref{fig:why}).

\begin{center}
\small
\begin{tabular}{@{}llll@{}}
\toprule
\textbf{Discipline} & \textbf{Interpretive field} & \textbf{Schema} & \textbf{Actual} \\
\midrule
Seismology & Instrument response \cite{stationxml} & Optional & \textbf{100\%} \\
Metrology & Measurement technique & Optional & 53.8\% \\
Meteorology & Sampling strategy \cite{wmdr} & Optional & 16.6\% \\
Volatile chemistry & Thermal programme & Optional & $\approx$0\% \\
\bottomrule
\end{tabular}
\end{center}

Every one of these fields is optional in its schema, and completeness ranges from
total to nil, so the schema does not explain the outcome. Neither does the
presence of a validator: the seismic instrument response is declared with
\texttt{minOccurs="0"}, has no validator enforcing it, and we measured 1{,}607
responses across 1{,}607 channels on live data centre holdings.

The rule that survives all four cases is that \textbf{a field is complete when the
primary consumer cannot complete the primary task without it}. A seismometer
reports counts, which are not a physical quantity; without the transfer function
there is no retrievable measurement. A chromatogram without its thermal programme
still opens. The programme is needed to \emph{compare}, and comparison is somebody
else's task, later.

Policy alone achieves nothing. The meteorological metadata standard is mandatory
by international agreement for all internationally exchanged observational data,
and across 179 operational stations and 1{,}444 deployments in ten territories,
sampling strategy is recorded on 16.6\%, spatial sampling resolution on 0.0\%, and
sample treatment on 6.6\% of which every instance states \texttt{unknown}. The
best-populated field is instrument height above the reference surface, at 62.1\%,
which is the field one cannot interpret a temperature without.

\section{Learnability of the derived corpus}

If collection were the binding constraint, a dataset assembled from the public
record should carry no usable signal. We tested this using 3{,}522 molecules with
113 binary odour labels, described by twenty character counts over a structure
string, fitted with closed-form ridge regression and evaluated by five-fold
cross-validation.

Mean AUC was \textbf{0.816} against a shuffled-label null of 0.481, with 45 of 51
labels above 0.70 and none at or below its own null. Training on that corpus and
testing on a separately curated corpus of 4{,}565 molecules gave a mean transfer
AUC of \textbf{0.772} with all 50 testable labels above 0.60. The strongest
classes were sulfurous, garlic, onion and alliaceous, consistent with a model
reading chemistry rather than an artefact.

This is a floor rather than a frontier. The features are the weakest structural
description available and any learned representation would improve on them. The
claim is only that a signal exists in publicly derived data, not that this
approaches published odour prediction.

One property of the evaluation is worth generalising. An aggregate check must fail
closed on an empty input. A verdict of the form \texttt{not X or mean(X) > t} is
satisfied by an empty $X$, so any join that silently returns nothing produces a
passing result rather than an error. The two corpora here key their tables by
different identifier systems, one by registry number and one by chemical
abstracts number, so the direct join does return nothing and the check does pass.

The pattern is not confined to this evaluation. It occurs in conformance layers
whose constraints match no node, and in completeness tables computed from a failed
network fetch, both of which report success in the same way and for the same
reason.

\section{The conformance ledger}
\label{sec:ledger}

Every repository we audited reports its holdings and none reports its defects. A
resource that publishes only what it has invites consumers to assume the rest.

We propose that a data resource publish a conformance ledger: a machine-readable,
versioned report of open findings against its own rules, with severity, count and
the condition that would clear each one. The count is then a tracked quantity
rather than an embarrassment.

Our released graph carries zero structural violations and \textbf{531 open
warnings}, every one of them stating that a percept recorded without a stimulus
concentration cannot be compared with another panel's. That rule is left firing
deliberately. Perceived quality changes with concentration and can reverse, so a
descriptor at an unstated concentration is a claim whose conditions are unknown,
and silencing the rule would conceal the resource's largest weakness. As
concentrations are added the count falls, which makes it a maintainable progress
metric.

This is not a new idea in every field. Crystallography has published structured
validation alerts with a formal reply mechanism for decades \cite{iucr}, and the
global legal entity register publishes the corroboration grade and lapse state of
every record \cite{gleif}.
It is absent from the scientific data repositories examined here.

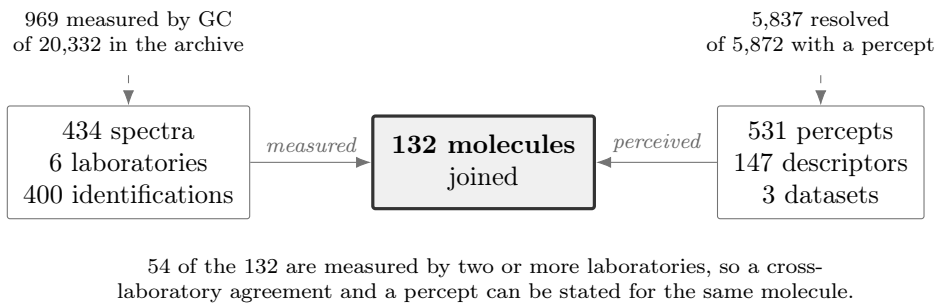
\begin{figure}[t]
\centering
\begin{tikzpicture}[
  node distance=7mm and 16mm,
  box/.style={draw=black!55, rounded corners=1pt, align=center, font=\small,
              inner sep=6pt, minimum height=11mm},
  hub/.style={draw=black!80, very thick, rounded corners=1pt, align=center,
              font=\small, inner sep=7pt, fill=black!6},
  lbl/.style={font=\scriptsize\itshape, text=black!60},
  ln/.style={-{Latex[length=2mm]}, draw=black!55}
]
\node[hub] (mol) {\textbf{132 molecules}\\joined};
\node[box, left=of mol] (meas) {434 spectra\\6 laboratories\\400 identifications};
\node[box, right=of mol] (perc) {531 percepts\\147 descriptors\\3 datasets};
\node[box, above=4mm of meas, draw=none, font=\scriptsize] (ms) {969 measured by GC\\of 20{,}332 in the archive};
\node[box, above=4mm of perc, draw=none, font=\scriptsize] (ps) {5{,}837 resolved\\of 5{,}872 with a percept};
\draw[ln] (meas) -- node[lbl, above]{measured} (mol);
\draw[ln] (perc) -- node[lbl, above]{perceived} (mol);
\draw[ln, dashed] (ms) -- (meas);
\draw[ln, dashed] (ps) -- (perc);
\node[below=5mm of mol, font=\scriptsize, text width=0.72\textwidth, align=center]
  {54 of the 132 are measured by two or more laboratories, so a cross-laboratory
   agreement and a percept can be stated for the same molecule.};
\end{tikzpicture}
\caption{The join an olfactory resource presupposes, at the scale the public record
supports. The two sides are not merged: a spectrum is an assertion by an
instrument and a percept an assertion by a panel, neither is evidence for the
other, and they meet at the molecule only, so either can be revised without
touching the other.}
\label{fig:bridge}
\end{figure}

\section{The released corpus}
\label{sec:dataset}

The measurements above are reproducible from public sources, but reproducing them
requires re-harvesting several repositories, one of which declined further
automated access during this work. We therefore release the derived graphs
themselves, expressed in the vocabulary described here, so that the results can be
inspected and queried without repeating the harvest.

\begin{center}
\small
\begin{tabular}{@{}lrl@{}}
\toprule
\textbf{Component} & \textbf{Triples} & \textbf{Contents} \\
\midrule
Derived comparisons & 33{,}115 & Cross-laboratory agreement from 137{,}030 spectra \\
Bridge & 10{,}514 & 132 molecules with measurement and percept \\
Recovered thermal programmes & 3{,}633 & 206 programmes, 529 ordered steps \\
Perception provenance & 444 & 53 datasets, 13 derived-field notices \\
Recovered reference series & 360 & 66 retention index series \\
Inferred descriptor relations & 290 & 29 surviving a permutation null \\
\midrule
\textbf{Released total} & \textbf{48{,}356} & \\
\addlinespace
\textit{Withheld: interlaboratory comparisons} & \textit{86{,}226} & \textit{4{,}082 degrees of equivalence, no licence declared} \\
\bottomrule
\end{tabular}
\end{center}

One component is withheld, and the reason is the paper's own argument applied to
itself. The interlaboratory comparison graph is the largest of the seven, and no
licence is declared anywhere on the source: we searched the captured page context
of all 443 retrieved identifiers and found no statement of terms. The publisher
also declined further automated access partway through a crawl its robots policy
permitted, and we stopped rather than continuing. Withholding it costs the deposit
roughly two thirds of its triples, and a project arguing that terms must be
recorded cannot redistribute 86{,}226 triples taken from a source that declares
none and asked us to stop. The harvesting and building scripts remain, so the
graph is reproducible by anyone the publisher permits, and every figure derived
from it in this paper is reported with the sampling caveat that the retrieved
slice is contiguous rather than random.

Three further properties of the release are deliberate and are not standard practice.

Every recovered value carries the sentence it was extracted from, in
\texttt{olf:recoveryEvidence}, so a reader can adjudicate any extraction by reading
rather than by rerunning. Every recovered value is also a dated assertion
attributed to the extracting pipeline rather than to the depositor, because an
extraction is a claim about what a text says and must not be laundered into a
statement its author made.

Every graph carries its own conformance result. The bridge validates with zero
structural violations and 531 open warnings, all of them stating that a percept
recorded without a stimulus concentration cannot be compared with another panel's.
That rule is left firing deliberately, as discussed in Section~\ref{sec:ledger}.

Records that map to no declared term are excluded and counted rather than assigned
an invented one. The bridge excludes 134{,}924 spectra on the ground that they
measure the molecule but not its volatility, and reports that exclusion rather than
absorbing it.

\section{Limitations}

The scalar rung is an upper bound, for the reason given in Section \ref{sec:chem}.

The comparability ladder is an analogy carried across domains. The logic of
graded, separately tested levels transfers; the factor-analytic apparatus that
establishes them in psychometrics does not, and none is used here.

Sixteen of 53 perception archives carry a compared vocabulary. Several of the
remainder are reference compilations rather than experiments. Term normalisation
merges spelling variants and not synonyms, making the reported overlap
conservative on spelling and optimistic on meaning.

The bridge treats a percept as a descriptor asserted for a molecule, not a rated
intensity at a stated concentration. Identifier resolution runs through one
registry, so 132 is a floor.

The separation between literatures reported in Section~\ref{sec:related} is
measured as co-occurrence of terminology in titles and abstracts, not as citation.
Two fields could cite each other's work without using each other's vocabulary, so
the figure is a lower bound on contact rather than a demonstration of isolation. A
citation-level analysis would be stronger and we have not performed one.

One data source declined further automated requests partway through a crawl its
robots policy permitted. We stopped rather than continuing, and the resulting
corpus of 56 comparisons and 4{,}082 degrees of equivalence over 186 institutes is
a contiguous slice of the identifier space rather than a sample. No proportion
should be computed from it.

\section{Data and code availability}

The released corpus is described in Section~\ref{sec:dataset} and is deposited at
\href{https://doi.org/10.5281/zenodo.22199924}{\texttt{10.5281/zenodo.22199924}},
with \texttt{10.5281/zenodo.22199923} resolving to all versions. The vocabulary
comprises seven OWL modules and six SHACL \cite{shacl} layers over an OWL 2 \cite{owl2} core, with 69 regression tests. All figures reported here are produced by numbered pipeline
scripts, and every headline number is computed twice, set-based and by query over
the graph, with the script exiting non-zero on disagreement. Derived graphs are
regenerable rather than committed where source licensing is unclear.


\begin{thebibliography}{99}
\small
\bibitem{meredith1993} W. Meredith. Measurement invariance, factor analysis and factorial invariance. \emph{Psychometrika}, 58:525--543, 1993.
\bibitem{kovats1958} E. Kov\'ats. Gas-chromatographische Charakterisierung organischer Verbindungen. \emph{Helvetica Chimica Acta}, 41:1915--1932, 1958.
\bibitem{vandendool1963} H. van den Dool and P. D. Kratz. A generalization of the retention index system including linear temperature programmed gas-liquid partition chromatography. \emph{Journal of Chromatography}, 11:463--471, 1963.
\bibitem{rohrschneider1966} L. Rohrschneider. Eine Methode zur Charakterisierung von gaschromatographischen Trennfl\"ussigkeiten. \emph{Journal of Chromatography}, 22:6--22, 1966.
\bibitem{mcreynolds1970} W. O. McReynolds. Characterization of some liquid phases. \emph{Journal of Chromatographic Science}, 8:685--691, 1970.
\bibitem{wang1992} Y. Wang and B. R. Kowalski. Calibration transfer and measurement stability of near-infrared spectrometers. \emph{Applied Spectroscopy}, 46(5):764--771, 1992.
\bibitem{fortier2017} I. Fortier, P. Raina, et al. Maelstrom Research guidelines for rigorous retrospective data harmonization. \emph{International Journal of Epidemiology}, 46(1):103--105, 2017.
\bibitem{wilson1927} E. B. Wilson. Probable inference, the law of succession, and statistical inference. \emph{Journal of the American Statistical Association}, 22:209--212, 1927.
\bibitem{horai2010} H. Horai, M. Arita, S. Kanaya, et al. MassBank: a public repository for sharing mass spectral data for life sciences. \emph{Journal of Mass Spectrometry}, 45:703--714, 2010.
\bibitem{haug2013} K. Haug et al. MetaboLights: an open-access general-purpose repository for metabolomics studies and associated meta-data. \emph{Nucleic Acids Research}, 41:D781--D786, 2013.
\bibitem{sud2016} M. Sud et al. Metabolomics Workbench: an international repository for metabolomics data and metadata, metabolite standards, protocols, tutorials and training, and analysis tools. \emph{Nucleic Acids Research}, 44:D463--D470, 2016.
\bibitem{hamel2024} E. A. Hamel et al. Pyrfume: a window to the world's olfactory data. \emph{Scientific Data}, 11:1220, 2024.
\bibitem{keller2016} A. Keller and L. B. Vosshall. Olfactory perception of chemically diverse molecules. \emph{BMC Neuroscience}, 17:55, 2016.
\bibitem{dravnieks1985} A. Dravnieks. \emph{Atlas of Odor Character Profiles}. ASTM DS61, 1985.
\bibitem{en13725} CEN. \emph{EN 13725: Stationary source emissions. Determination of odour concentration by dynamic olfactometry and odour emission rate}. 2022.
\bibitem{stationxml} International Federation of Digital Seismograph Networks. \emph{FDSN StationXML schema, version 1.2}. \url{https://www.fdsn.org/xml/station/}
\bibitem{wmdr} World Meteorological Organization. \emph{WIGOS Metadata Representation, version 1.0}. \url{https://schemas.wmo.int/wmdr/}
\bibitem{cipmmra} Bureau International des Poids et Mesures. \emph{Measurement comparisons in the CIPM MRA: guidelines for organizing, participating and reporting}. CIPM MRA-G-11 version 1.2, 2025.
\bibitem{shacl} World Wide Web Consortium. \emph{Shapes Constraint Language (SHACL)}. W3C Recommendation, 2017.
\bibitem{owl2} World Wide Web Consortium. \emph{OWL 2 Web Ontology Language Document Overview}. W3C Recommendation, 2012.
\bibitem{gleif} Global Legal Entity Identifier Foundation. \emph{LEI data and ISO 17442}. \url{https://www.gleif.org/}
\bibitem{iucr} International Union of Crystallography. \emph{checkCIF and the Validation Reply Form}. \url{https://checkcif.iucr.org/}
\bibitem{openalex} J. Priem, H. Piwowar and R. Orr. OpenAlex: a fully-open index of scholarly works, authors, venues, institutions, and concepts. \emph{arXiv:2205.01833}, 2022.
\bibitem{pubchem} S. Kim et al. PubChem in 2021: new data content and improved web interfaces. \emph{Nucleic Acids Research}, 49:D1388--D1395, 2021.
\end{thebibliography}
\end{document}